\documentstyle{article}
\newcommand{\be}[1]{\begin{equation}\label{#1}}
\newcommand{\ee}{\end{equation}}
\newcommand{\rf}[1]{(\ref{#1})}
\newcommand{\gorder}{\mbox{ \small $^{>}_{\sim}$ }}
\newcommand{\lorder}{\mbox{ \small $^{<}_{\sim}$ }}
\begin{document}
\title{\large \bf
MULTIDIMENSIONAL QUANTUM COSMOLOGY: QUANTUM WORMHOLES, THIRD
QUANTIZATION, INFLATION FROM "NOTHING", etc
\thanks
 {To appear in the proceedings of the 6th Quantum Gravity Seminar,
 Moscow, 12-19 June 1995 }}
\author{\large \bf A. ZHUK}
\date{}
\maketitle
\medskip
\begin{center}
Department of Theoretical Physics\\ University of Odessa, 2 Petra
Velikogo St.\\ 270100 Odessa, Ukraine\\
\end{center}
\medskip
\medskip
\medskip
{\bf Abstract}\\

A multidimensional cosmological model with space-time consisting of
$n\ (n\geq2)$ Einstein spaces $M_i$ is investigated in the presence
of a cosmological constant $\Lambda$ and $m\ (m\geq1)$ homogeneous
minimally coupled scalar fields as a matter source. Classes of the
models integrable at classical as well as quantum levels are found.
These classes are equivalent to each other. Quantum wormhole
solutions are obtained for them and the procedure of the third
quantization is performed. An inflationary universe arising from
classically forbidden Euclidean region is investigated for a model
with a cosmological constant.

\large
%
%
\newpage
\section{Introduction}

We believe that for description of quantum gravitational processes
at high energies the multidimensional approach is more adequate.
Modern theories of unified physical interactions use ideas of
hidden (or extra) dimensions. In order to study different phenomena
at early stage of the universe one should use these theories or at
any rate models keeping their main ideas. In any case,
multidimensional models have to explain the observed four
dimensionality of space-time at present time. This is realized in
models where one or a number of internal spaces are compact and
contracted to Planckian scales during the evolution of the universe
(dynamical compactification) or the symmetry between the external
and internal dimensions is broken from the very beginning and the
internal spaces are static and compactified at Planck's length
(spontaneous compactification). A further possibility is given in
the quantum theory where only the external spaces can be created by
tunneling while the internal dimensions may be hidden because they
stay behind a potential barrier.

   Of special interest are exact solutions because they can be
used for a detailed study of the evolution of our space, of the
compactification of the internal spaces and of the behavior of
matter fields.

One of the most natural mulidimensional cosmological models (MCM)
generalizing the Friedmann-Robertson-Walker (FRW) universe is given
by a toy model with the topology $R\times M_1\times\ldots\times M_n$
where $M_i\ (i=1,\ldots ,n)$ denotes Einstein space. One of these
spaces, say $M_1$, describes the external space but all others are
internal spaces. The gauge covariant form of the Wheeler-De Witt
(WDW) equation \cite{1,2} for a model with this topology was
proposed in \cite{3} and some integrable models were investigated
in \cite{4}-\cite{6}.

In the recent paper we consider a general model with a cosmological
constant $\Lambda$ and $m\ (m\geq 1)$ homogeneous minimally coupled
scalar fields $\varphi^{(a)}\ (a=1,\ldots,m)$ with potentials
$U^{(a)}(\varphi^{(a)})$. We show that some integrable models
considered before \cite{4}-\cite{6} and some new ones are
equivalent to each other. Among solutions of the WDW equations for
these models there are ones which describe tunneling universes, in
particular, birth of universes from classically forbidden Euclidean
region (birth from "nothing" \cite{7}). Quantum wormholes \cite{8}
representing a special class of solutions of the WDW equation are
constructed for considered models also. Full sets of the orthonormal
solutions in these models give possibility to perform third
quantization \cite{9} and to get a spectrum of created universes.
An inflationary universe arising due to quantum tunneling is
investigated for a model with a cosmological constant. Parameters
of the model which ensure inflation of the external space and
dynamical compactification of internal spaces are found. In
particular, the dimension of internal spaces should be $d > 40$. It
is shown that the tunneling from "nothing" of this universe is
strongly suppressed because of very large spatial volume of the
arisen universe.
%
\section{General description of the model}

The metric of the model
\be{2.1}
g=-\exp{[2\gamma(\tau)]}d\tau\otimes d\tau +
\sum_{i=1}^n\exp{[2\beta^i(\tau)]}g_{(i)}
\ee
is defined on the manifold
$$
M= R\times M_1\times\ldots\times M_n\ ,
$$
where the manifold $M_i$ with the metric $g_{(i)}$ is an Einstein
space of dimension $d_i$, i.e.
\be{2.2}
R_{m_i n_i}[g_{(i)}]=\lambda^i g_{(i)m_i n_i}
\ee
$$
i=1,\ldots,n;\ n\geq 2.
$$
The total dimension of the space-time $M$ is
$D=1+\sum^{n}_{i=1}d_i$. This describes the case where the
topology of a factorized space-time manifold is assumed from the
very beginning and compactification of internal spaces is described
as shrinking to or freezing on the Planck scale.

Here we investigate the general model with cosmological constant
$\Lambda$ and $m\ (m\geq 1)$ non-interacting homogeneous minimally
coupled scalar fields $\varphi^{(a)}\ (a=1,\ldots,m)$ with
potentials $U^{(a)}(\varphi^{(a)})$. The action of the model is
adopted in the following form:
\be{2.3}
S=\frac{1}{2\kappa^2}\int d^D x\sqrt{|g|}\left\{
R[g]-2\Lambda\right\} + S_{\varphi} + S_{GH}\ ,
\ee
where $R[g]$ is the scalar curvature of the metric \rf{2.1} and
$\kappa^2$ is D-dimensional gravitational constant. $S_{GH}$ is the
standard Gibbons-Hawking boundary term \cite{10}. $S_{\varphi}=
\sum_{a=1}^{m}S_{\varphi}^{(a)}$ is the action of $m$
non-interacting minimally coupled homogeneous scalar fields
\be{2.4}
S_{\varphi}^{(a)}=\int d^D x\sqrt{|g|}\left[
-\frac{1}{2}g^{MN}\partial_{M}\varphi^{(a)}\partial_{N}\varphi^{(a)}
- U^{(a)}(\varphi^{(a)})
\right]\ .
\ee
For the metric \rf{2.1} the action \rf{2.3} reads
\be{2.5a}
S=\mu\int d\tau L
\ee
with the Lagrangian $L$ being
\be{2.5}
L=\frac{1}{2} e^{-\gamma+\gamma_0}\left(
G_{ij}\dot{\beta}^i \dot{\beta}^j +
\kappa^2 \sum_{a=1}^m\left(\dot{\varphi}^{(a)}\right)^2\right)
- V\ .
\ee
Here $\gamma_0=\sum_{i=1}^n d_i\beta^i$ and the overdot denotes
differentiation with respect to the time $\tau$. The components of
the minisuperspace read \cite{3}
\be{2.6}
G_{ij}=d_i\delta_{ij}-d_i d_j\ ,
\ee
and the potential is given by
\be{2.7}
V=e^{\gamma+\gamma_0}
\left(
-\frac{1}{2}\sum_{i=1}^n \theta_i
e^{-2\beta^i} +
\kappa^2\sum^m_{a=1}U^{(a)}(\varphi^{(a)}) + \Lambda
\right)\ ,
\ee
where $\theta_i=\lambda^i d_i$. If the $M_i$ are spaces of constant
curvature, then $\theta_i$ may be normalized in such a way that
$\theta_i=k_i d_i(d_i - 1),\ k_i=\pm 1, 0$. The parameter
$\mu=\prod^n_{i=1} V_i/ \kappa^2$ where $V_i$ is the volume of
$M_i$ and we may put $\mu=1$ \cite{11}.

The constraint equation reads
\be{2.8}
-\frac{\partial L}{\partial \gamma}=
\frac{1}{2} e^{-\gamma+\gamma_0}
\left(
G_{ij}\dot{\beta}^i \dot{\beta}^j + 
\kappa^2 \sum_{a=1}^m\left(\dot{\varphi}^{(a)}\right)^2
\right)+ V = 0\ .
\ee

The minisuperspace metric $G=G_{ij}d\beta^i\otimes d\beta^j$ may be
diagonalized in different coordinate systems. In the present paper
we use two of them. In the first one the minisuperspace metric
reads \cite{4,5,12}
\be{2.9}
G=-dv^0\otimes dv^0 + \sum^{n-1}_{i=1}dv^i\otimes dv^i ,
\ee
where
$$
q_1v^0=(d_1-1)\beta^1+\sum_{i=2}^n d_i\beta^i ,
$$
$$
q_1v^1=\left[(D-2)/(d_1\Sigma_2)\right]^{1/2}\sum_{i=2}^n d_i\beta^i ,
$$
\be{2.10}
q_1v^i=\left[(d_1-1)d_i/d_1\Sigma_i\Sigma_{i+1})\right]^{1/2}
\sum_{j=i+1}^n d_j(\beta^j-\beta^i),\ i=2,\ldots,n-1 .
\ee
Here we used the notations $q_1=\left[(d_1-1)/d_1\right]^{1/2}$
and $\Sigma_i=\sum_{j=i}^n d_j$. In the second coordinate system
\be{2.11}
G=-dz^0\otimes dz^0 + \sum_{i=1}^{n-1}dz^i\otimes dz^i \ ,
\ee
where \cite{3,6,13}
$$
z^0=q^{-1}_2\sum_{i=1}^n d_i\beta^i ,
$$
\be{2.12}
z^i=\left[d_i/ \Sigma_i\Sigma_{i+1})\right]^{1/2}
\sum_{j=i+1}^n d_j(\beta^j-\beta^i) ,\ 
i=1,\ldots,n-1.
\ee
Here $q_2=\left[(D-1)/(D-2)\right]^{1/2}$. The spatial volume of
the universe is proportional to $v=\prod_{i=1}^n a_{i}^{d_i}$ where
scale factors $a_i=\exp{\beta^i}$. In the coordinates \rf{2.12}
it takes the following form
\be{2.13}
v=\prod_{i=1}^n a_{i}^{d_i} = \exp{(q_2 z^0)} .
\ee
%
%

\section{Wheeler-De Witt equations. Integrable cosmologies}

\setcounter{equation}{0}

At the quantum level the constraint \rf{2.8} is modified into
the WDW equation \cite{1,2}. Now, we consider classes of
cosmological models which are integrable at classical as well as
quantum levels, and we show the equivalence of these models.

\subsection{Models with one non-Ricci-flat factor space}

In the present chapter we consider the integrable case of a MCM where
only one of the factor space $M_i$, say $M_1$, is not Ricci-flat:
$\theta_1\neq 0,\  \theta_i=0,\  i=2,\ldots,n$. Using the coordinates
\rf{2.10} we set for free scalar fields (it is clear that in the
case of free scalar fields it is sufficient to take $m=1$) the
following form of the constraint \rf{2.8}
\be{3.1}
-\left({{\dot{v}}^0}\right)^2+\sum_{i=1}^{n-1}\left({\dot{v}}^i\right)^2+
{\dot{\varphi}}^2- \theta_1 e^{2q_1 v^0} = 0\ ,
\ee
where we used the harmonic time gauge $\gamma=\gamma_0$ and the
scalar field $\varphi^{(1)}\equiv \varphi$ is redefined: $\kappa\varphi\ 
\rightarrow\ \varphi$. 
The WDW equation in this case reads \cite{4,5,12,14}
\be{3.2}
\left( -\frac{\partial}{\partial v^0}\frac{\partial}{\partial v^0}+
\sum_{i=1}^{n-1} \frac{\partial}{\partial v^i}\frac{\partial}{\partial v^i}+
\frac{\partial^2}{\partial \varphi^2}+ \theta_1 e^{2q_1
v^0}\right)\Psi= 0\ .
\ee
It is easy to obtain solutions of this equation by separation of variables
\be{3.3}
\Psi(v)= e^{i{\bf pv}}\Phi(v^0)\ ,
\ee
where ${\bf p}= (p^1,\ldots,p^n)$ is a constant vector,
${\bf v}= (v^1,\ldots,v^{n-1},v^n=\varphi),\  {\bf pv}= 
\sum_{i=1}^n p_iv^i$ and $p_i=p^i$. The substitution of \rf{3.3}
into \rf{3.2} gives
\be{3.4}
\left[ -\frac{1}{2}\left( \frac{d}{dv^0}\right)^2+ 
\frac{1}{2}\theta_1 e^{2q_1 v^0}\right]\Phi= \varepsilon \Phi\ ,
\ee
where
\be{3.5}
\varepsilon= \frac{1}{2}\sum_{i=1}^n (p^i)^2 \ .
\ee
\subsection{Models with cosmological constant}

Here we consider the integrable case of a MCM with all Ricci-flat
factor-spaces: $\theta_i=0\ (i=1,\ldots,n)$ in the presence of the
cosmological constant $\Lambda$ and free scalar field as a matter source.
Using the coordinates \rf{2.12} and the harmonic time gauge we get for
\rf{2.8} 
\be{3.6}
-(\dot{z}^0)^2+ \sum_{i=1}^{n-1} (\dot{z}^i)^2+ \dot{\varphi}^2+
2\Lambda e^{2q_2z^0}= 0\ ,
\ee
which is modified into the WDW equation \cite{3,6,13}
\be{3.7}
\left( -\frac{\partial}{\partial z^0}\frac{\partial}{\partial z^0}+
\sum_{i=1}^{n-1} \frac{\partial}{\partial z^i}\frac{\partial}{\partial z^i}+
\frac{\partial^2}{\partial \varphi^2}- 2\Lambda e^{2q_2
z^0}\right)\Psi= 0\ .
\ee
We are seeking the solution of \rf{3.7} in the form
\be{3.8}
\Psi(z)= e^{i{\bf pz}}\Phi(z^0)\ .
\ee
where ${\bf p}= (p^1,\ldots,p^n)$ and ${\bf z}=
(z^0,\ldots,z^{n-1},z^n=\varphi)$. 
 $\Phi(z^0)$ satisfies the equation
\be{3.9}
\left[ -\frac{1}{2}\left( \frac{d}{dz^0}\right)^2- 
\Lambda e^{2q_2 z^0}\right]\Phi= \varepsilon \Phi
\ee
with $\varepsilon$ defined by the relation \rf{3.5}.

It is easy to see that the models of the subsections 3.1 and 3.2 are
equivalent to each other with an accuracy to the evident substitutions:
$$
v^i\ \leftrightarrow \ z^i,\  i=0,\ldots,n-1,
$$
$$
\frac{1}{2}\theta_1\ \leftrightarrow \ -\Lambda,
$$
\be{3.10}
q_1\ \leftrightarrow \ q_2\ .
\ee
Thus, to investigate the quantum behavior of these models it is sufficient to
consider one of them, e.g. that of subsection 3.2 only.
\subsection{Exact scalar field cosmologies}

Here we consider a special class of the integrable MCM with $m\ (m\geq 1)$
scalar fields. The action of these models is given by the relations
\rf{2.3}, \rf{2.4} where $\Lambda= 0$. The Lagrangian \rf{2.5} for
these models reads
\newpage
$$
L_s=\frac{1}{2} e^{-\gamma+\gamma_0}\left(
G_{ij}\dot{\beta}^i \dot{\beta}^j + 
\kappa^2 \sum_{a=1}^m\left(\dot{\varphi}^{(a)}\right)^2\right)+
$$
\be{3.11}
e^{\gamma+\gamma_0}\left(\frac{1}{2}\sum_{i=1}^n \lambda^i d_i
e^{-2\beta^i}- \kappa^2\sum_{a=1}^m U^{(a)}(\varphi^{(a)})\right).
\ee

The energy-momentum tensor of a matter with the action
$S=\int d^Dx \sqrt{|g|}L$ is defined by
\be{3.12}
T_{ik}=-2\frac{\partial L}{\partial g^{ik}}+ g_{ik}L .
\ee
Using this formula we get the non-zero components of the scalar
field $\varphi^{(a)}$ energy-momentum tensor:
\be{3.13}
{T^{(a)}}^0_0=
-\frac{1}{2}e^{2\gamma}\left(\dot{\varphi}^{(a)}\right)^2- 
U^{(a)}(\varphi^{(a)})\equiv -\rho^{(a)} ,
\ee
\be{3.14}
{T^{(a)}}^{m_i}_{m_i}=
\frac{1}{2}e^{2\gamma}\left(\dot{\varphi}^{(a)}\right)^2- 
U^{(a)}(\varphi^{(a)})\equiv P^{(a)} ,\ i=1,\ldots,n , 
\ee
where we introduced the energy density $\rho^{(a)}$ and the
pressure $P^{(a)}$ which correspond to the scalar field
$\varphi^{(a)}$. Now we suppose that these quantities are connected
by the equation of state 
\be{3.15}
P^{(a)}= \left(\alpha^{(a)}-1\right)\rho^{(a)} ,
\ee
where $\alpha^{(a)}= const,\ a=1,\ldots,m$.

It is not difficult to prove that these models are equivalent to
the cosmological models in the presence of $m$-component perfect
fluid with the energy-momentum tensor
\be{3.16}
T^M_N= \sum_{a=1}^m {T^{(a)}}^M_N ,
\ee
\be{3.17}
{T^{(a)}}^M_N= diag\left(-\rho^{(a)}(\tau),
P^{(a)}(\tau)\delta^{m_1}_{k_1},\ldots,{P^{(a)}(\tau)}\delta^{m_n}_{k_n}
\right) ,
\ee
where for any $a$-th component of perfect fluid the pressure and the
energy density are connected by the equation of state \rf{3.15} and
conservation equations are imposed on each component separately:
\be{3.18}
{T^{(a)}}^M_{N;M}= 0.
\ee
The non-trivial conservation equations \rf{3.18} are
\be{3.19}
{\dot{\rho}}^{(a)}+ {\dot{\gamma}}_0\left(\rho^{(a)}+ P^{(a)}\right)= 0,\ 
a=1,\ldots,m ,
\ee
which results in 
\be{3.20}
{\rho}^{(a)}= A^{(a)}\exp{(-\alpha^{(a)}\gamma_0)}=
A^{(a)}v^{-\alpha^{(a)}}\ ,\ a=1,\ldots,m ,
\ee
where $A^{(a)}= const$ and the spatial volume $v$ is defined by \rf{2.13}.

To prove the proposition about equivalence between these models we note first
that the Klein-Gordon equations
\be{3.21}
\frac{\partial}{\partial\tau}
\left(e^{-\gamma+\gamma_0}\ \frac{\partial\varphi^{(a)}}{\partial\tau}\right)+
e^{\gamma+\gamma_0}\ \frac{\partial U^{(a)}}{\partial \varphi^{(a)}}= 0,\ 
a=1,\ldots,m ,
\ee
following from the Lagrangian \rf{3.11} are equivalent to the conservation
equations \rf{3.19} if the relations \rf{3.13} and \rf{3.14} have
place. Second, using the relations \rf{3.13} -- \rf{3.15}  and
\rf{3.20} we obtain
\be{3.22}
\frac{\partial L_s}{\partial\gamma}=\frac{\partial L_{\rho}}{\partial\gamma}
\ee
and
\be{3.23}
\frac{d}{d\tau} \frac{\partial L_s}{\partial\dot{\beta}^i}
-\frac{\partial L_{s}}{\partial \beta^i}=
\frac{d}{d\tau} \frac{\partial L_{\rho}}{\partial\dot{\beta}^i}-
\frac{\partial L_{\rho}}{\partial \beta^i},\ 
i=1,\ldots,n ,
\ee
where the lagrangian $L_s$ is given by \rf{3.11} and the 
effective Lagrangian $L_{\rho}$ corresponding to the models with the
energy-momentum tensor \rf{3.16} -- \rf{3.18} is \cite{15a,15,16}
\be{3.24}
L_{\rho}=\frac{1}{2} e^{-\gamma+\gamma_0}
G_{ij}\dot{\beta}^i \dot{\beta}^j + 
e^{\gamma+\gamma_0}\left(\frac{1}{2}\sum_{i=1}^n \lambda^i d_i
e^{-2\beta^i}- \kappa^2\sum_{a=1}^m \rho^{(a)}\right) ,
\ee
with $\rho^{(a)}$ defined by \rf{3.20}.

To get integrable models we consider the Ricci-flat case 
$(\lambda^i=0,i=1,\ldots,n)$. Then, in the $z$-coordinates \rf{2.12}
the Lagrangian and the constraint read respectively
\be{3.25}
L_{\rho}= -\frac{1}{2}\left( \dot{z}^0\right)^2+
\frac{1}{2}\left( \dot{z}^i\right)^2-
\kappa^2\sum_{a=1}^m A^{(a)}\exp\left[(2-\alpha^{(a)})q_2 z^0\right]
\ee
and
\be{3.26}
 -\frac{1}{2}\left( \dot{z}^0\right)^2+
\frac{1}{2}\left( \dot{z}^i\right)^2+
\kappa^2\sum_{a=1}^m A^{(a)}\exp\left[(2-\alpha^{(a)})q_2
z^0\right]\ ,
\ee
where we use the harmonic time gauge $\gamma=\gamma_0$. The equations
of the motion
\be{3.27}
\ddot{z}^i= 0,\ i=1,\ldots,n-1
\ee
have the first integrals
\be{3.28}
\dot{z}^i= p^i.
\ee
The constraint \rf{3.26} can be rewritten as follows
\be{3.29}
\dot{v}= \pm \sqrt{2}q_2 v\left[\varepsilon+ \kappa^2 v^2
\sum_{a=1}^m \rho^{(a)}(v) \right]\ ,
\ee
where the parameter $\varepsilon $ is defined by the relation similar to
\rf{3.5}. Using the relation \rf{3.29} and \rf{3.13} it is not difficult to
get $\varphi^{(a)}$ as a function of the spatial volume \cite{17}
\be{3.30}
\varphi^{(a)}= \pm\frac{\sqrt{\alpha^{(a)}/2}}{q_2}
\int\frac{\sqrt{\rho^{(a)}(v)}dv}{\left[\varepsilon+ 
\kappa^2 v^2 \sum_{a=1}^m \rho^{(a)}(v)\right]^{1/2}}+
\varphi_0^{(a)}\ .
\ee
Inverting this expression we can find the spatial volume as a function 
of the scalar field $\varphi^{(a)}:\ v=v(\varphi^{(a)})$ and 
consequently a dependence of the energy density $\rho^{(a)}$ on the
scalar field $\varphi^{(a)}:\ \rho^{(a)}=\rho^{(a)}(\varphi^{(a)})$.
To reconstruct the potentials $U^{(a)}$ we can write them with the
help of the relations \rf{3.13} -- \rf{3.15} in the form
\be{3.31}
U^{(a)}(\varphi^{(a)})= \frac{1}{2}(2-\alpha^{(a)})
\rho^{(a)}(\varphi^{(a)}),\ \ a=1,\ldots,m,
\ee
where
\be{3.32}
\rho^{(a)}(\varphi{(a)})= A^{(a)}\left[v(\varphi^{(a)})
 \right]^{-\alpha^{(a)}}\ .
\ee

If $\alpha^{(a)}=0$, then $\varphi^{(a)}, U^{(a)}$ and $\rho^{(a)}$
are constant. This scalar field is equivalent to the cosmological
constant $\Lambda\equiv \kappa^2 U^{(a)}$ with the equation of
state $P^{(a)}= -\rho^{(a)} $. For $\alpha^{(a)}=2 $ we have
$U^{(a)}\equiv 0$. The scalar field $\varphi^{(a)}$ in this case is
equivalent to the perfect fluid with the ultra-stiff equation of
state $P^{(a)}= \rho^{(a)}$.

It is clear that it is hardly possible to integrate models with an
arbitrary set of exponents in the relations \rf{3.25} and \rf{3.26}.
Therefore, we consider now the integrable case of the two-component
$(m=2)$ scalar field with an arbitrary
$\alpha^{(1)}\equiv\alpha\neq0,\ 2$ and $\alpha^{(2)}=2$ which
corresponds to the the two-component perfect fluid with the energy
density 
\be{3.33}
\rho= \rho^{(1)}+ \rho^{(2)}= A^{(1)}v^{-\alpha}+ A^{(2)}v^{-2}\ .
\ee
The potentials $U^{(a)}$ in this case read \cite{17}\ 
$U^{(2)}\equiv0$ and
$$
U^{(1)}(\varphi^{(1)})=
\frac{(2-\alpha)\Lambda}{2\kappa^2}\exp\left[\mp\sqrt{2\alpha}\kappa
q_2(\varphi^{(1)}-\varphi^{(1)}_0) \right],
\ E=0,\ \Lambda>0, \ 
$$
$$
= U^{(1)}_0
\left[ \sinh{\left( 
\frac{\kappa q_2 (2-\alpha)}{\sqrt{2\alpha}} 
(\varphi^{(1)}-\varphi^{(1)}_0) 
\right)}
\right]
^{-2\alpha/(2-\alpha)},
\ E>0,\ \Lambda>0, \ 
$$
$$
= - U^{(1)}_0\left[ \sin{\left( 
\frac{\kappa q_2 (2-\alpha)}{\sqrt{2\alpha}} 
(\varphi^{(1)}-\varphi^{(1)}_0)
\right)}
\right]^{-2\alpha/(2-\alpha)},
\ E>0,\ \Lambda<0, \ 
$$
\be{3.34}
= U^{(1)}_0\left[ \cosh{\left( 
\frac{\kappa q_2(2-\alpha)}{\sqrt{2\alpha}} 
(\varphi^{(1)}-\varphi^{(1)}_0)
\right)}
\right]^{-2\alpha/(2-\alpha)},
\ E<0,\ \Lambda<0\ ,
\ee
where $\Lambda= \kappa^2 A^{(1)}$ and 
\be{3.35}
E= \varepsilon+ \kappa^2 A^{(2)}\ .
\ee

As usual, at quantum level the constraints $\partial
L_s/\partial\gamma=0$ and $\partial
L_{\rho}/\partial\gamma=0$ are modified into the WDW equations.
It is more simple to investigate a quantum behavior of the geometry
in these models studying the WDW
equation which was obtained from the effective Lagrangian
\rf{3.25}. For the integrable two-component model \rf{3.33} we get
\be{3.36}
\left( -\frac{\partial}{\partial z^0}\frac{\partial}{\partial z^0}+
\sum_{i=1}^{n-1} \frac{\partial}{\partial z^i}\frac{\partial}{\partial z^i}-
2\kappa^2 A^{(2)}- 2\kappa^2 
A^{(1)}e^{(2-\alpha)q_2z^0}\right)\Psi= 0\ .
\ee

Separating the variables in the form
\be{3.37}
\Psi(z)= e^{i{\bf pz}}\Phi(z^0),
\ee
where
${\bf p}= (p^1,\ldots,p^{n-1})$ is a constant vector and 
${\bf z}=(z^1,\ldots,z^{n-1})$, we get for $\Phi(z^0)$ the equation
\be{3.38}
\left[-\frac{1}{2}\left(\frac{d}{dz^0} \right)^2- \kappa^2
A^{(1)}e^{(2-\alpha)q_2 z^0} \right]\Phi= E\Phi
\ee
with $E$ defined by \rf{3.35}.

It is easy to see again that the models of the subsections 3.2 and 3.3
are equivalent to each other also with an accuracy to the
substitutions: 
$$
\varepsilon\ \leftrightarrow \ E,
$$
$$
\Lambda\ \leftrightarrow \ \kappa^2 A^{(1)},
$$
\be{3.39}
2q_2\ \leftrightarrow \ (2-\alpha)q_2.
\ee
%
\section{Quantum wormholes}
\setcounter{equation}{0}

Using the equivalency between these three models we can consider
one of them only to investigate quantum behavior of the universe.
Let us consider the second of them. Simple analysis of the equation
\rf{3.9} shows \cite{6} that quantum behavior depends strongly on
the signs of $\Lambda$ and $\varepsilon$. For example, if
$\Lambda>0$ then for $\varepsilon\geq0$ the Lorentzian regions
exist only. For $\varepsilon<0$ both regions, the Lorentzian as
well as Euclidean one exist. In this case quantum transitions with
topology changes take place (tunneling universes or birth from
"nothing").  If $\Lambda <0$ then for $\varepsilon\leq0$ the
Euclidean region exists only. For $\varepsilon>0$ both regions,
the Lorentzian as well as Euclidean one exist. In this case quantum
transitions with topology changes take place (quantum wormholes).

Let us consider the latter case in more detail. Solving \rf{3.9},
we get
\be{4.1}
\Phi(z^0)= B_{i\sqrt{2\varepsilon}/q_2}\left(
\sqrt{-2\Lambda}q_2^{-1}\exp{(q_2 z^0)}
\right),
\ee
where
$\sqrt{2\varepsilon}/q_2= |{\bf p}|/q_2$ and $B= I, K $ are
modified Bessel functions. The general solution of the equation
\rf{3.7} has the following form:
\be{4.2}
\Psi(z)= \sum_{B=I,K} \int d^n  p\ C_{B}({\bf p})
\exp{(i{\bf pz})} B_{i|{\bf p}|/q_2}
\left(
\sqrt{-2\Lambda}q_2^{-1}\exp{(q_2z^0)}
\right),
\ee
where
functions $C_B\ (B=I,K)$ belong to an appropriate class.

Quantum wormholes represent a special class of solutions of the WDW
equation with the following boundary conditions \cite{8}:

(i) \ \ \ the wave function is exponentially damped for large
spatial geometry,

(ii) \ \ \ the wave function is regular when the spatial geometry
degenerates. 

\noindent We restrict our consideration to real values of $p_i,\
(i=1,\ldots,n)$ . This corresponds to real geometries in Lorentzian
region. In this case we have $\varepsilon\geq 0$.

If $\Lambda>0$ the wave function \rf{3.8} $\Psi=\exp{(i{\bf
pz})}\ \Phi(z^0)$ with $\Phi(z^0)$ defined by \rf{4.1} is not
exponentially damped when the spatial volume $v\ \rightarrow \ \infty$,
i.e. the condition (i) for quantum wormholes is not satisfied. It
oscillates and may be interpreted as corresponding to the classical
Lorentzian solutions.

For $\Lambda<0$ the wave function \rf{3.8} is exponentially damped
for large $v$ only when $B=K$ in \rf{4.1}. But in this case the
function $\Phi$ oscillates an infinite number of times when $v\
\rightarrow\ 0$. Thus, the condition (ii) is not satisfied.The
wave function describes the transition between Lorentzian and
Euclidean regions.

The function
\be{4.3}
\Psi_{\bf p}(z)= \exp{(i{\bf pz})}\ K_{i|{\bf p}|/q_2}
\left(
\sqrt{-2\Lambda}q_2^{-1}\exp{(q_2 z^0)}
\right)
\ee
may be used for constructing quantum wormhole solutions. We
consider the superposition of singular solutions
\be{4.4}
\Psi_{\lambda,{\bf n}}(z)= \frac{1}{\pi}\int_{-\infty}^{+\infty}
dk \Psi_{q_2 k{\bf n}}(z)\exp{(-ik\lambda)},
\ee
where $\lambda\in R,\ {\bf n}$ is a unit vector $({\bf n}^2=1)$ and
the quantum number $k$ is connected with the quantum number
$\varepsilon=\frac{1}{2}{|\bf p|}^2$ by the formula $2\varepsilon=
q_2^2 k^2$. The calculation gives \cite{6,13}
\be{4.5}
\Psi_{\lambda,{\bf n}}(z)= \exp{\left(
-\frac{\sqrt{-2\Lambda}}{q_2}\exp{(q_2 z^0)}\cosh{(\lambda-q_2{\bf
zn})} 
\right)}\ .
\ee
It is not difficult to verify that the formula \rf{4.5} leads to
solutions of the WDW equation \rf{3.7}, satisfying the quantum
wormholes boundary conditions. Similar quantum wormholes for the
model of subsection 3.1 were obtained in \cite{5,12}.

These results are the straightforward generalization of the
discussion in \cite{18} -- \cite{20} to the multidimensional case.
Therefore, the set of wave fnctions $\Psi_{\bf p}$ and
$\Psi_{\lambda ,{\bf n}}$ are spanning the same space of physical
states and are both bases of the Hilbert space of the model in the
corresponding representation. The connection between these bases
$\Psi_{\bf p}$ and $\Psi_{\lambda ,{\bf n}}$ is given by the equation
\rf{4.4}.

The function
\be{4.6}
\Psi_{m,{\bf n}}= H_m(x^0) H_m(x^1)
\exp{
\left\{
-\frac{1}{2}\left[(x^0)^2+ (x^1)^2\right]
\right\}\ ,
}
\ee
where $H_m$ are the Hermite polynomials and
\be{4.7}
x^0= (2/q_2)^{1/2}(-2\Lambda)^{1/4}\exp(q_2
z^0/2)\cosh{(\frac{1}{2}q_2{\bf zn})},
\ee
\be{4.8}
x^1= (2/q_1)^{1/2}(-2\Lambda)^{1/4}\exp(q_2
z^0/2)\sinh{(\frac{1}{2}q_2{\bf zn})},
\ee
$m= 0,1,\ldots$ are also solutions of the WDW equation with the
quantum wormholes boundary conditions. Solutions of such type are
called discrete spectrum quantum wormholes \cite{5,8,12,13}, \cite{18} --
\cite{20} and form a discrete basis for the Hilbert space of the
system. 
%
%
\section{Third quantization}
\setcounter{equation}{0}

The WDW equations \rf{3.2}, \rf{3.7} and \rf{3.36} are similar to
the scalar field equation in the curved space-time. These equations
are gauge covariant (conformal covariant) \cite{3}. It is not
difficult to show \cite{4,21} that the minisuperspace metric $G$
\rf{2.6} is conformally equivalent to the Milne metric and for a
special gauge the WDW equations \rf{3.2}, \rf{3.7} and \rf{3.36} 
coincide with a field equation for a scalar field conformally
coupled to a Milne space-time.

By analogy with the quantum field theory it might be worth-while to
perform the second quantization of the universe wave function
$\Psi$ expanding it on the creation and annihilation operators.
The WDW equation itself is a result of the quantization of a
geometry and matter. Thus, the procedure of the wave function
$\Psi$ quantization is called the \ {\it third quantization}
\cite{9}\ . Similar to the quantum scalar field theory in the curved 
space-time we can expect that the vacuum state in a third quantized
theory is unstable and creation of particles (in our case,
universes) from the initial vacuum state takes place. To perform
the procedure of scalar field quantization on the time-dependent
gravitational field background we should determine the vacuum
state. This is quite a problem. Since there is no global timelike
Klling vector and, hence, there is no global vacuum state, it is
only possible to define different vacuum states which, in general,
are not equivalent to each other and have different physical
nature. By analogy with this, in the third quantization procedure
we have a similar situation. Different Fock spaces constructed from
the exact solutions of the WDW equations are not equivalent to each
other. It is natural to define the initial vacuum state with
respect to the orthonormal set of mode solutions which are positive
frequency modes in the limit of vanishing spatial volume: $v\
\rightarrow \ 0$ \cite{22}. As a result, the birth of particles
(universes) from "nothing" may have place where "nothing" is
defined above initial vacuum state.

Let us consider now the model of subsection 3.2 with $\Lambda>0$,
which corresponds to a scalar field with the positive square of
mass. It is not difficult to find two complete set of modes
\cite{4,21} 
\be{5.1}
\Psi_{\bf p}= \frac{1}{(2\pi)^{n/2}}
\frac{\sqrt{\pi}}{\left[2q_2
\sinh{(\pi\sqrt{2\varepsilon}/q_2)}\right]^{1/2}} 
e^{i{\bf pz}} J_{-i\sqrt{2\varepsilon}/q_2}
\left(
\frac{\sqrt{2\Lambda}}{q_2}e^{qz^0}
\right)
\ee
and
\be{5.2}
\hat{\Psi}_{\bf p}= \frac{1}{(2\pi)^{n/2}}
\frac{\sqrt{\pi}}{2\sqrt{q_2}}
\exp{(
\pi\sqrt{\varepsilon/2}/q_2)} 
e^{i{\bf pz}} H^{(2)}_{i\sqrt{2\varepsilon}/q_2}
\left(
\frac{\sqrt{2\Lambda}}{q_2}e^{qz^0}
\right)\ ,
\ee
which are orthonormal:
\be{5.3}
\left(
\Psi_{\bf p},\ \hat{\Psi}_{\bf p\prime}
\right)=
-i\int_{z^0=const}\Psi_{\bf p}\
\stackrel{\leftrightarrow}{\partial_{z^0}}\ \hat{\Psi}_{\bf
p\prime}^{*} d{\bf z}= \delta({\bf p-p\prime}).
\ee
The modes \rf{5.1} are excited states above the Hartle-Hawking
vacuum state \cite{23} with $\varepsilon=0$ \cite{6}. As both sets,
\rf{5.1} and \rf{5.2} are complete, they are releted to each other
by the Bogolubov transformation 
\be{5.4}
\Psi_{\bf p}= \alpha_{\varepsilon}\hat{\Psi}_{\bf p}+
\beta_{\varepsilon}\hat{\Psi}_{\bf p}^{*},
\ee
where the Bogolubov coefficients are
\be{5.5}
\alpha_{\varepsilon}=
\left[
\frac{\exp{(\pi\sqrt{2\varepsilon}/q_2)}}
{2\sinh{(\pi\sqrt{2\varepsilon}/q_2)}}
\right]^{1/2}
\ee
and
\be{5.6}
\beta_{\varepsilon}=
\left[
\frac{\exp{(-\pi\sqrt{2\varepsilon}/q_2)}}
{2\sinh{(\pi\sqrt{2\varepsilon}/q_2)}}
\right]^{1/2}\ .
\ee
The coefficients $\beta_{\varepsilon}$ are not equal to zero. Thus,
two Fock spaces constructed with the help of the modes
$\Psi_{\bf p}$ and $\hat{\Psi}_{\bf p}$ are not equivalent and we
have two different third quantized vacuum states (voids): $|0>$ and
$|\hat{0}>$. The modes \rf{5.1} have the asymptote
\be{5.7}
\Psi_{\bf p}\ \sim\ \exp{\left[ i({\bf
pz}-\sqrt{2\varepsilon}z^0)\right]}, \ v\rightarrow 0\ ,
\ee
which are positive frequency modes with respect to the conformal
"time" $z^0$. Thus, the vacuum $|0>$ which is defined with respect
to these modes is connected to the minisuperspace conformal Killing
vector $\partial_{z^0}$. The modes \rf{5.1} are no longer positive
frequency ones under $v\ \rightarrow \ \infty$. In this limit the
modes \rf{5.2} have the asymptote 
\be{5.8}
\hat{\Psi}_{\bf p}\ \sim\ \exp{\left[ i({\bf
py}-\sqrt{2\Lambda}y^0)\right]}, \ v\rightarrow \infty\ ,
\ee
where
\be{5.9}
y^0= q_2^{-1}\exp{(q_2 z^0)}\ ;\ y^i=z^i,\ i=1,\ldots,n.
\ee
The modes \rf{5.2} in this limit are positive frequency ones with
respect to the "time" $y^0$.

Since the vacuum states $|0>$ and $|\hat{0}>$ are not equivalent, 
the birth of the universes from "nothing" may have place, where
"nothing" is the vacuum state $|0>$. If $|0>$ is the initial state
when $v \ \rightarrow \ 0$, then an observer defined with respect
to the vacuum state $|\hat{0}>$ will detect in the limit $v \ \rightarrow
\ \infty$
\be{5.10}
n_{\varepsilon}= |\beta_{\varepsilon}|^2=
\left[
\exp{(2\pi\sqrt{2\varepsilon}/q_2)}- 1
\right]^{-1}
\ee
universes in mode ${\bf p}$ (we remind that $2\varepsilon= {\bf
p}^2$). This is precisely Planck spectrum for radiation at
temperature $T= q_2/2\pi$.

Now we consider the model with $\Lambda <0$. It is not difficult
to see that we can get the WDW equation \rf{3.7} from the action
\be{5.11}
S= \frac{1}{2}\int d^{n+1} z\Psi\hat{H}\Psi,
\ee
which coincides with the action for a scalar field in the Minkowski
space-time with the potential 
\be{5.12}
V(\Psi)= \frac{M^2}{2}{\Psi}^2,
\ee
where
\be{5.13}
M^2 (z)= 2\Lambda\exp{(2q_2z^0)}.
\ee
If $\Lambda <0$,then $M^2 <0$ and this model has an unstable vacuum
state. The spectrum of energy is unbounded from below. The theory
is well defined if we add the self-interaction term. Then
\be{5.14}
V(\Psi)= -\frac{\nu^2}{2}\Psi^2+ \frac{\lambda}{4}\Psi^4\ ,
\ee
where we define
\be{5.15}
M^2\equiv -\nu^2= -2|\Lambda|\exp{(2q_2z^0)}\ .
\ee
The minimum of the potential \rf{5.14} has place at
\be{5.16}
\Psi_0= \pm\frac{\nu}{\sqrt{\lambda}}= \pm 
\sqrt{\frac{2|\Lambda|}{\lambda}}\exp{(q_2 z^0)}.
\ee
It follows from this expression that symmetry breaking takes place
dynamically, because
\be{5.17}
\Psi_0\ \rightarrow \ 0\ , if \ \ v\ \rightarrow \ 0.
\ee
The depth of wells at minima is:\ \ \ \ \ \ $V(\Psi=\Psi_0) = 
-\nu^4/(4\lambda) =  -(\Lambda^2/\lambda)\exp{(4q_2 z^0)}$. The
square of mass of the field $\Psi$ excitations after symmetry
breaking becomes positive:
\be{5.18}
m^2(\Psi=\Psi_0)= \left.\frac{d^2 V}{d\Psi^2}\right|_{\Psi=\Psi_0}= 2\nu^2.
\ee
Let us consider now the field $\bar{\Psi}= \Psi-\Psi_0$ which
describes oscillations near minima of the potential $V(\Psi)$. This
field satisfies the equation
\be{5.19}
\left( -\frac{\partial}{\partial z^0}\frac{\partial}{\partial z^0}+
\sum_{i=1}^{n} \frac{\partial}{\partial z^i}\frac{\partial}{\partial z^i}-
2\nu^2\right)\bar{\Psi}=
j(z^0)+ 3\lambda\Psi_0 \bar{\Psi}^2+ \lambda\bar{\Psi}^3\ ,
\ee
where the source $j$ is
\be{5.20}
j(z^0)= \frac{\partial}{\partial z^0}\frac{\partial}{\partial
z^0}\Psi_0= \pm q_2^2 \sqrt{\frac{2|\Lambda|}{\lambda}}\exp{(q_2 z^0)}.
\ee
As it follows from the relation \rf{5.18}, the field $\bar{\Psi}$
has positive square of mass which depends on the "time" $z^0$ .
Thus, in linear approximation and without the source term the birth
{}from "nothing" takes place as for the case $\Lambda >0$. We should
make in the formulas \rf{5.1} and \rf{5.2} the only replacement:
$\Lambda \rightarrow 2|\Lambda|$. Presence of the source term in the
equation \rf{5.19} leads to an additional universes production. The
source term has its origin in the dependence of the classical
minimum $\Psi_0$ on "time" (see Eq. \rf{5.20}).

Presence of the interaction terms $\sim \bar{\Psi^2}$ and
$\bar{\Psi^3}$ in the relation \rf{5.19} (respectively, $\sim
\bar{\Psi}^3$ and $\bar{\Psi}^4$ in the potential $V(\bar{\Psi})$) gives
the possibility to consider the processes with the topology
alteration. For example, the cubic term in the potential is
analogous to the interaction term which arises naturally in the
string theory. This term describes the fission of the universe into
two or the fusion of two universes into one.

It is important to note that the third quantization may have an
influence on choice of the topology of models. If we demand the
renormalizability of third quantized theory, then, by analogy with
the scalar field theory with self-interaction, it follows that its
dimension should be equal or less four \cite{24}. In our case it
means that we should take models with $n \leq 3$, i.e. in the
models without scalar field $\varphi$ we can take at most four
factor-spaces $M_i$ and in the presence of scalar field we can
consider at most three factor-spaces.
%
%
\section{Inflation from "nothing"}
\setcounter{equation}{0}

Inflationary models are very popular now in cosmology because they
explain why our universe is homogeneous, isotropic and almost
spatially flat \cite{25}. So, it might be worth-while to get
inflationary models in multidimensional cosmology also. However,
contrary to usual 4-dimensional space-time cosmologies, in the
multidimensional case we should solve two problems simultaneously.
Namely, it is necessary to get inflation of our external space and
compactification of internal dimensions near Planck length
$L_{Pl}\sim 10^{-33} \ cm$ to make them unobservable at present
time. 

Another interesting hypothesis consists in the proposal that
iflationary universe arose by quantum tunneling from classically
forbidden Euclidean region. This process is called the birth from
"nothing" \cite{7} as in the previous chapter 5, but its nature is
quite different to former one.

In present chapter we investigate the multidimensional inflationary
universe  which arose from "nothing" by quantum tunneling process.
It is clear that in the harmonic time gauge the solutions of the
constraints \rf{3.1}, \rf{3.6} and \rf{3.26} and the equation of the
form \rf{3.27} are equivalent to each other, but inverting them we
get quite different behavior of the scale factors $a_i=
exp{\beta^i}$ in these models. Thus, the analysis of classical
behavior of the universe for each model should be performed
separately. 

Here we investigate the model of the subsection 3.2 with the
cosmological constant $\Lambda >0$. As it follows from the chapter
4, in this case the quantum tunneling takes place if $\varepsilon <
0$. As we demand the reality of metric in the Lorentzian region,
the condition $\varepsilon <0$ takes place for imaginary scalar field
$\varphi$ only \cite{6}.

In the harmonic time gauge the solution of the constraint \rf{3.6}
reads 
\be{6.1}
v= \exp{(q_2 z^0)}=
\frac{\sqrt{|\varepsilon|/\Lambda}}{\cos{(q_2\sqrt{2|\varepsilon|}\tau)}}\
\ ,\ 
\ |\tau|\ \leq \ \frac{\pi/2}{q_2\sqrt{2|\varepsilon|}}\ ,
\ee
with the turning point at minimum of the spatial volume
\be{6.2}
v_{min}= \sqrt{|\varepsilon|/\Lambda}\equiv v_t\ .
\ee
The analytic continuation $\tau_L \rightarrow -i\tau_E$ gives the
solution in the Euclidean region
\be{6.3}
v=
\frac{\sqrt{|\varepsilon|/\Lambda}}{\cosh{(q_2\sqrt{2|\varepsilon|}\tau)}}\
, \ -\infty\ <\tau\ <+\infty
\ee
with the turning point at maximum: $v_{max}= v_t$.

In synchronous system $(\gamma=0)$ the scale factors read \cite{6}
\be{6.4}
a_i= A_i\left[
\cosh{(\frac{t}{T})}
\right]^{\sigma}
\left[
f(\frac{t}{2T})
\right]^{\sigma_i},\ i=1,\ldots,n,
\ee
where $t$ is the proper time, $\sigma= 1/(D-1)$ and $T= \left[
(D-2)/2\Lambda(D-1)\right]^{1/2}$. Here
\be{6.5}
f(x)= \exp{\left[ -2\arctan{e^{-2x}}\right]}
\ee
is smooth monotonically increasing function with the asymptotes:
$f(x)\ \rightarrow \ \exp{(-\pi)}$ as $x\rightarrow -\infty$, 
$f(x)\ \rightarrow \ 1$ as $x\rightarrow +\infty$, and at zero:
$f(0)= \exp{(-\pi/2)}$. The parameters $\sigma_i$ satisfy the
relation 
\be{6.6}
\sum^n_{i=1} d_i\sigma^i= 0
\ee
and
\be{6.7}
\sum^n_{i=1} d_i\sigma^2_i+ \sigma_{n+1}^2= \frac{D-2}{D-1}\ .
\ee

The spatial volume reads
\be{6.8}
v= \left(
\prod^n_{i=1} A_i^{ d_i}\right)\cosh{\frac{t}{T}}
\ee
and has its minimum at $t=0$. It is not difficult to verify that
\be{6.9}
\prod^n_{i=1} A_i^{ d_i}= \sqrt{|\varepsilon|/\Lambda}.
\ee
The scale factors $a_i$ have its minima at
\be{6.10}
\frac{t_{(0)i}}{T}= \mbox{ arsinh }\frac{\sigma_i}{\sigma}=
-\ln{\left[
\frac{\sigma_i}{\sigma}+ \sqrt{\left(\frac{\sigma_i}{\sigma}\right)^2+ 1}
\right]},\ i=1,\ldots,n,
\ee
{}from which it follows that $sign\ t_{(0)i}= -sign\ \sigma_i$.

We suppose now that the universe arose tunneling from the Euclidean
region and from the turning point $t=0$ (see Eq. \rf{6.8}) its
behavior can be described by classical equations. For simplicity,
we consider the model with 
two factor spaces $(n=2)$ where one of them (say $M_1$) is our
external space. The generalization to the case $n> 2$ is
straightforward. We suppose also that after birth the external
space $M_1$ monotonically expands. Thus, it follows from the
equations \rf{6.6} and \rf{6.10} that $t_{(0)1}< 0\ (\sigma_1> 0)$
and $t_{(0)2}> 0\ (\sigma_2< 0)$. Let all dimensions at the moment
of the universe creation from "nothing" have equal rights:
\be{6.11}
a_1(t=0)= a_2(t=0)= 10^x\ L_{Pl},
\ee
where $2\leq x\leq 3$. If we take $x>3$ then the probability of the
birth becomes too small because of too large spatial volume. If
$x<2$ then the scale factor $a_2$ goes to $L_{Pl}$ too fast and it
is not sufficient time for inflation of the scale factor $a_1$.
{}From the relations \rf{6.4} and \rf{6.11} we get 
\be{6.12}
A_i=  10^x\exp{\frac{\pi}{2}\sigma_i}\ ,\ i=1,2.
\ee
Using these relations we find for the scale factor $a_1$ that
\cite{26} 
\be{6.13}
a_1\approx 10^x\exp{\frac{\pi}{2}\sigma_1},
\ee
if
\be{6.14}
4\ \lorder \ t/T \ \ll D-1.
\ee

To solve the flatness and horizon problems the scale factor $a_1$
during inflation should expand in $10^{30}$ times \cite{27}.
Thus, 
\be{6.15}
\frac{\pi}{2}\sigma_1\ \gorder \ 70.
\ee
It gives the lower boundary for the parameter $\sigma_1$.
If the size of $M_1$ to the end of inflation is approximately equal
to the size 
of observable at the present time universe, i.e. $\sim 10^{28}\ cm$, then
\be{6.16}
\frac{\pi}{2}\sigma_1\ \approx 140.
\ee
For the parameter $\sigma_2$ we get
\be{6.17}
\sigma_2 \ \lorder \ -\frac{d_1}{d_2}\frac{140}{\pi}.
\ee
It can be easily seen that within the limits $140/\pi\leq\sigma_1\leq
280/\pi$ we have for the position of the minima of $a_2$
\be{6.18}
t_{0(2)}/T\ \approx \ 6,
\ee
if $d_2\gg d_1= 3$. Here we consider the model where the space $M_2$ shrinks 
at the end of the inflation to its minimum size near Planck length, i.e.
\be{6.19}
t_{0(2)} \ = \ t^{*}
\ee
and
\be{6.20}
a_2(t_{0(2)}) \ \approx \ L_{Pl}.
\ee
Thus, for the scale factor $a_2$ we get
\be{6.21}
a_2(t_{0(2)}) \ \approx \ L_{Pl}\ \approx \ 
10^x \exp{(\frac{\pi}{2}\sigma_2)} \ \lorder \ 10^x
\exp{(-\frac{70d_1}{d_2})}, 
\ee
which gives an estimate
\be{6.22}
\frac{70d_1}{d_2}\ \lorder\ x\ln{10}.
\ee
For example, if $d_1=3,\ \frac{\pi}{2}\sigma_1=70$ and $2\leq x\leq
3$ we get respectively
\be{6.23}
45\ \geq\  d_2 \ \geq 30.
\ee
In general, we find that to ensure the inflation of the external space the
dimension of the internal space should be $d> 40$ in accordance with the paper
\cite{28}.

After inflation the external space $M_1$ should have a power-law expansion 
and the internal space $M_2$ should remain frozen near Planck scale. The 
transition to
such stage can be performed if the cosmological constant $\Lambda$ goes very
fast to zero. As a result we have the Kasner-like solution \cite{29}:
$a_i=a_{(0)i}t^{\alpha_i}, \varphi= \ln{t^{\alpha_{n+1}}}+const$, where
$\sum^n_{i=1} d_i \alpha_i= 1$ and $\sum^n_{i=1} d_i {\alpha_i}^2=
1-{\alpha_{n+1}}^2$. In particular, the solution with the freezed internal
spaces exists when $\alpha_i=0\ (i=2,\ldots\,n)$. In this case for the extrenal
space we get $\alpha_1=1/d_1$. Thus, the factor-space $M_1$ expands as a FRW
universe filled with ultra-stiff matter (for $d_1=3$).

Now we consider the probability of the birth from "nothing" for the
inflationary universe. The amplitude of transition between the states with 
zero
spatial volume $v=0$ at the moment $\tau_i$ and some value of $v$ at $\tau_f$
is given by the path integral 
\be{6.24}
<v,\tau_f|\oslash,\tau_i>= \int [dg] [d\varphi ] e^{iS_L},
\ee
where $S_L$ is a Lorentzian action and the path integral is taken over all
trajectories between points $v=0$ and $v$. In our case the action $S_L$ is
given by the relation \rf{2.5a} (for all Ricci-flat factor spaces 
and one-component free scalar field ($m=1$)) and we consider the
transition between 
"points" with $v=0$ and the classical turning point $v_t=
\sqrt{|\varepsilon|/\Lambda}$. To make the oscillating integral \rf{6.24}
convergent it is necessary to perform the Vick rotation to Euclidean time:
$\tau_L\ \rightarrow\ -i\tau_E$. The probability between points
$v=0$ and $v_t$ 
is proportional to square of modulus of amplitude:
\be{6.25}
P\sim |<v_t,\tau_f|\oslash,\tau_i>|^2.
\ee
In semiclassical limit
\be{6.26}
<v_t,\tau_f|\oslash,\tau_i> \sim e^{-S_E}\ ,
\ee
where the Euclidean action for our model
\be{6.27}
S_E= \frac{1}{2\kappa^2}\int^{\tau_f}_{\tau_i} d\tau\left[
-(\dot{z}^0)^2+ 2|\varepsilon|+ 2\Lambda e^{2q_2 z^0}
\right]- \left.\frac{1}{2\kappa^2}\frac{\dot{v}}{v}\right|_{\tau_i}=
\ee
$$
=2\frac{\Lambda}{\kappa^2}\int_{\tau_i}^{\tau_f} d\tau e^{2q_2
z^0}- \left.\frac{1}{2\kappa^2}\frac{\dot{v}}{v}\right|_{\tau_i}
$$
is calculated on classical solutions of the Euclidean field equations
(instantons) interpolating between the vanishing geometry $v=0$ and the 
turning
point $v_t$.

Classical solutions of the Euclidean field equations in our model are given by
the relation \rf{6.3} which shows that $\tau_i=-\infty\ (v=0)$ and $\tau_f=0\
(v=v_t)$. Substituting this relation into \rf{6.27} we get \cite{26}
\be{6.28}
S_E= C\left[
8\pi\prod^n_{i=2} a^{d_i}_{(c)i}
\right]^{-1}\sqrt{|\varepsilon|},
\ee
where
\be{6.29}
C= \left(\frac{\sqrt{2}}{q_2}-\frac{q_2}{\sqrt{2}}\right)
\ee
and we took into account that $D$-dimensional gravitation constant $\kappa^2$
is connected with the Newton constant $G_N$ by the relation
\be{6.30}
\kappa^2= 8\pi G_N \prod_{i=2}^n a_{(c)i}^{d_i},
\ee
where $a_{(c)i}$ are the scale factors of freezed internal spaces
(in the formula \rf{6.28} we put $G_N=1$). The parameter $C>0$ for
$D>3$ and the presence of the boundary term in the action \rf{6.27}
does not change the sign of $S_E$. Let us estimate $S_E$ for the
two-component ($n=2$) inflationary model. It follows from the
relations \rf{6.6}, \rf{6.9} and \rf{6.12} that
\be{6.31}
v_t= \sqrt{|\varepsilon|/\Lambda}= A_1^{d_1}A_2^{d_2}=
10^{x(d_1+d_2)}.
\ee
Then, with the help of the estimate \rf{6.22} we  get
\be{6.32}
\sqrt{|\varepsilon|}= \Lambda^{1/2} 10^{x(d_1+d_2)} \gorder
\Lambda^{1/2} 10^{d_1(x+70/\ln{10})}.
\ee
Thus,
\be{6.33}
S_E= C\left[8\pi a_{(c)2}^{d_2}\right]^{-1}\sqrt{|\varepsilon|}
\gorder \frac{C}{8\pi}\Lambda^{1/2} 10^{d_1(x+70/\ln{10})},
\ee
where $a_{(c)2}\approx L_{Pl}$ (see Eq. \rf{6.20}). 
The quantum birth of universe will be not suppressed if $S_E\lorder
 1$. It follows from the formula \rf{6.33} that it takes place for 
$\Lambda\lorder 10^{-124} cm^{-2}$.
This quantity is much less than possible value of the cosmological 
constant in the observable universe $\Lambda_0\lorder 10^{-57}
cm^{-2}$. Thus, for realistic theories $S_E\gg 1$
and, consequently, the quantum birth of this system is strongly
suppressed. The reason of it consists in the large number of the
internal dimensions, as it follows from the relation \rf{6.31}.

In some of papers (see, e.g. \cite{27}) it was suggested that,
instead of the standard Euclidean rotation $\tau_L \rightarrow
-i\tau_E$, the action \rf{6.27} should be obtained by rotation in
the opposite sense, $\tau_L \rightarrow +i\tau_E$. However, we get
in this case non-physical result that the probability of the birth
of the universe is proportional to the volume of the arisen
universe. 

%
%

%

%

\begin{thebibliography}{99}

\bibitem{1}  J.A.Wheeler, Geometrodynamics (Academic, New York, 1962).

\bibitem{2}  B.C.De Witt, {\it Phys.Rev.}  \ {\bf 160} (1967) 1113.

\bibitem{3} V.D.Ivashchuk, V.N.Melnikov and A.I.Zhuk, {\it Nuovo Cimento} 
\ {\bf B140} (1989) 575.

\bibitem{4}  A.Zhuk, {\it Class.Quant.Grav.} \ {\bf  9 } (1992) 2029.

\bibitem{5}  A.Zhuk, {\it Phys.Rev.} \ {\bf  D45 } (1992) 1192.

\bibitem{6} U.Bleyer, V.D.Ivashchuk, V.N.Melnikov and A.Zhuk, {\it Nucl.Phys.}
\ {\bf B429 } (1994) 177.

\bibitem{7} P.I.Fomin, {\it Dokl.Akad.Nauk Ukr.SSR} \ {\bf 9A } (1975) 831;\\
A.Vilenkin, {\it Phys.Rev.}\ {\bf D27 } (1983) 2848, \ {\bf D50 }
(1994) 2581;\\ V.A. Rubakov, {\it Phys.Lett.}\ {\bf B148 } (1984)
280;\\ A.Linde, {\it Lett.Nuovo Cimento}\ {\bf 39 } (1984) 401;\\
Ya.B.Zeldovich and A.A.Starobinsky, {\it Sov.Astron.Lett.}\ {\bf 10 }
(1984) 135;\\ A.O.Barvinsky and A.Yu.Kamenshchik, {\it
Phys.Rev.}\ {\bf D50 } (1994) 5093. 

\bibitem{8}  S.W.Hawking and D.N.Page, {\it Phys.Rev.}\ {\bf D42 } (1990) 2665.

\bibitem{9} V.A.Rubakov, {\it Phys.Lett.}\ {\bf B214 } (1988) 503;\\ 
S.Giddings and
A.Strominger, {\it Nucl.Phys.}\ {\bf B321 } (1989) 481;\\ A.A.Kirillov, 
{\it Pis'ma
Zh.Eksp.Teor.Fiz.}\ {\bf 55 } (1992) 541.

\bibitem{10}  G.W.Gibbons and S.W.Hawking, {\it Phys.Rev.}\ {\bf D15 } (1977) 
2752.

\bibitem{11}  U.Bleyer and A.Zhuk, {\it Gravitation and Cosmology}\ {\bf 1 } 
(1995) 37.

\bibitem{12}  A.Zhuk, {\it Sov.J.Nucl.Phys.}\ {\bf 55 } (1992) 149.

\bibitem{13} V.D.Ivashchuk and V.N.Melnikov, {\it Teor.Mat.Fiz.}\ {\bf 98 } 
(1994) 312. 

\bibitem{14} U.Bleyer and A.Zhuk, {\it Gravitation and Cosmology}\ {\bf 1 } 
(1995) 106. 

\bibitem{15a}  U.Bleyer, D.-E.Liebscher, H.-J.Schmidt and A.Zhuk,
{\it Wissenschaftliche Zeitschrift} (Jena) \ {\bf 39} (1990) 22.

\bibitem{15}  V.D.Ivashchuk and V.N.Melnikov, {\it Int.J.Mod.Phys.}\ {\bf D3 } 
(1994) N4;\\
V.R.Gavrilov, V.D.Ivashchuk and V.N.Melnikov, Multidimensional Cosmology
with multicomponent perfect fluid and Toda lattices, Preprint gr-qc/9407019.

\bibitem{16}  V.D.Ivashchuk and V.N.Melnikov, Billiard representation for
multidimensional cosmology with multicomponent perfect fluid near the
singularity, Preprint gr-qc/9407028.

\bibitem{17}  A.Zhuk, Integrable scalar field multidimensional cosmologies,
(submitted to {\it Class.Quant.Grav.}, 1995).

\bibitem{18} L.M.Campbell and L.J.Garay, {\it Phys.Lett.}\ {\bf
B254 } (1991) 49. 

\bibitem{19}  L.J.Garay, {\it Phys.Rev.}\ {\bf D48 } (1993) 1710.

\bibitem{20}  G.A.Mena Marugan, {\it Phys.Rev.}\ {\bf D50 } (1994) 3923.

\bibitem{21}  A.Zhuk, {\it Sov.J.Nucl.Phys.}\ {\bf 58 } (1995) N11.

\bibitem{22}  Y.Peleg, {\it Class.Quant.Grav.}\ {\bf 8 } (1991) 827.

\bibitem{23}  J.B.Hartle and S.W.Hawking, {\it Phys.Rev.}\ {\bf D28 } 
(1983) 2960.

\bibitem{24}  J.Zinn-Justin, Quantum Field Theory and Critical Phenomena
(Claredon Press, Oxford, 1989).

\bibitem{25} A.D.Linde, {\it Phys.Rev.}\ {\bf D45 } (1994) 748;\\ A.Linde, 
D.Linde
and A.Mezhlumian, {\it Phys.Rev.}\ {\bf D49 } (1994) 1783.

\bibitem{26}  A.Zhuk, Inflation from ''nothing'' in multidimensional
cosmology \\ (to appear in {\it Sov.J.Nucl.Phys.},\ {\bf 59 } (1996)).

\bibitem{27}  A.Linde, {\it Rep.Prog.Phys.}\ {\bf 47 } (1984) 925.

\bibitem{28}  R.Abbot, S.Barr and S.Ellis, {\it Phys.Rev.}\ {\bf D30 } 
(1984) 720.

\bibitem{29}  U.Bleyer and A.Zhuk, Kasner-like, inflationary and steady
state solutions in multidimensional cosmology, Preprint of Astrophysical
Institute of Potsdam, AIP 95-03.

\end{thebibliography}
\end{document}